\documentclass[twocolumn,aps,showpacs,prl,amsmath,amssymb,floatfix,superscriptaddress]{revtex4}
\usepackage{color}
\usepackage{graphicx}
\usepackage{dcolumn}
\usepackage{bm}
\usepackage{array}
\usepackage{float}
\usepackage{supertabular}
\usepackage{longtable}
\usepackage{mathrsfs}
\usepackage{txfonts}
\usepackage{wasysym}

\begin{document}

\title{Anyonic Liquids in Nearly Saturated Spin Chains}

\author{Armin Rahmani}
\affiliation{
Theoretical Division, T-4 and CNLS, Los Alamos National Laboratory, Los Alamos, New Mexico 87545, USA} 

\author{Adrian E. Feiguin}
\affiliation{
Department of Physics, Northeastern University, Boston, Massachusetts 02115, USA}

\author{Cristian D. Batista}
\affiliation{
Theoretical Division, T-4 and CNLS, Los Alamos National Laboratory, Los Alamos, New Mexico 87545, USA} 

\date{\today}
\pacs{75.10.Jm, 71.10.Pm, 05.30.Pr}

\begin{abstract}
Most Heisenberg-like spin chains  flow to a universal free-fermion fixed point near the magnetic-field induced saturation point. Here we show that an exotic fixed point, characterized by two species of low-energy excitations with mutual anyonic  statistics, may also emerge in such spin chains if the dispersion relation has two minima. By using bosonization, two-magnon exact calculations,  and numerical density-matrix-renormalization-group, we demonstrate the existence of this anyonic-liquid fixed point in an XXZ spin chain with up to second neighbor interactions. We also identify a range of microscopic parameters, which support this phase.
\end{abstract}
 
\maketitle

Magnetic-field induced saturation of quantum magnets is one of the most widely studied quantum critical points (QCP) of nature: magnets with axial symmetry along the field axis become fully polarized at a critical field value. In two and three spatial dimensions, the corresponding QCP that separates the fully and partially polarized states belongs to the ``Bose-Einstein condensate'' (BEC) universality class.~\cite{Zapf14,Giamarchi08, Batyev84, Affleck90,Giamarchi99} The magnets can be treated as a dilute gas of bosons in the vicinity of the QCP by mapping the spins that are antiparallel to the field into hard-core bosons. In contrast, in most {\it one-dimensional ($d=1$) models} studied thus far, the weakly-interacting quasiparticles  near the field-induced QCP have \textit{fermionic} statistics ~\cite{Sachdev}. Here we demonstrate that a much richer spectrum of QCPs, including novel \textit{anyonic} liquids, may emerge in nearly saturated axially symmetric spin chains.

The essential ingredient  is magnetic frustration, which can provide natural realizations of single-particle dispersions with degenerate minima at multiple wave vectors ${\bf Q}$~\cite{note0}. Such single-particle dispersions do not change the universality class of the BEC QCP in $d>1$, but can give rise to multi-${\bf Q}$ condensates~\cite{Nikuni95,Nikuni00,Veillette05,Griset11} such as long-range ordered magnetic vortex crystals~\cite{Kamiya14,Marmorini14}. In contrast, long-range order is suppressed in $d=1$ due to strong quantum fluctuations. In this case, a Jordan-Wigner  (J-W) transformation~\cite{Jordan28,Batista01} allows us to describe the magnet as a dilute gas of interacting fermions near the  QCP. The Pauli exclusion principle  renders all fermion-fermion interactions irrelevant (in a renormalization-group sense), resulting in a free-fermion fixed point with a single-minimum dispersion relation~\cite{Sachdev}. The central question addressed in this paper is the fate of the $d=1$ QCP  when magnetic frustration generates a  dispersion relation with two degenerate minima.  
\begin{figure}
\vspace{0.3cm}
 \includegraphics[width =8 cm]{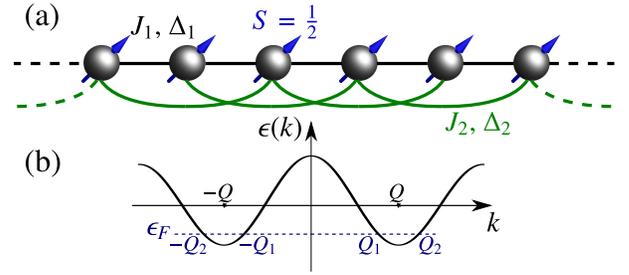}
\caption{The dispersion of Eq.~\eqref{eq:H_0} with two minima at $\pm Q$. The Fermi points are at $\pm Q_i$, $i=1,2$ with corresponding Fermi velocities $v_i$. }
 \label{fig:1}
\end{figure}

We show that frustration can stabilize a novel \textit{anyonic-liquid} near the field-induced QCP of spin chains. This result extends the classification of QCPs for saturated quantum magnets from simple theories of free bosons ($d>1$), and free fermions ($d=1$), to an exotic line of QCPs with  emergent Abelian anyonic statistics that interpolate between these two fixed points. Our anyonic-liquid consists of two species of quasiparticles originating from the two degenerate minima  (with two species of anyons, inversion symmetry breaking is not necessary and we consider on models with inversion symmetry~\cite{Hao09,Keilmann10}). Quasiparticles of different species do not interact with each other yet their commutation relations imply that they are Abelian anyons as opposed to simple bosons or fermions. In fact, similar theories of $d=1$ Abelian anyons~\cite{note1} have been envisioned in the field-theory literature through abstract flux attachment to free { bosonic theories~\cite{Schulz98,Kundu98,Kundu99,Pham00,Pham00b,Girardeau06,Batchelor06,Calabrese07,Patu07,Hao08,delCampo08,Batchelor08,Gils08,Finch13}.} However, no experimentally relevant microscopic models have been shown to support such anyonic liquids. By combining bosonization, renormalization-group arguments and numerical density-matrix renormalization group (DMRG) computations~\cite{White92,White93}, we  provide an experimentally relevant realization for these  elusive anyonic liquids in the context of frustrated magnetism. Moreover, we propose experimental signatures, which should facilitate their observation.

The corresponding  XXZ Hamiltonian~\cite{Chubukov91,White96,Vekua07,Hikihara08,Heidrich-Meisner09,Kolezhuk12,Shyiko13}, 
\begin{equation}\label{eq:xxz}
H=\sum_{j;a=1,2}\left[{J_a\over 2}\left(S_j^+ S^-_{j+a}+S^+_{j+a}S^-_j\right)+\Delta_a J_a \left(S^z_j S_{j+a}^z-{1\over 4}\right)\right],
\end{equation}
is illustrated in Fig.~\ref{fig:1}(a). It includes up to second-neighbor exchange interactions and a Zeeman term which allows to tune $S_T^z=\sum_j S^z_j$ with an external magnetic field $B_z$ ($S^z_T$ is conserved because $[H,S^z_T]=0$).  For a possible physical realization in a bilayer zigzag ladder, see Refs.~\cite{Batista09,Batista12}.

After a J-W transformation, $S^-_j=c_j \exp\left(-i \pi \sum_{k<j} n_k\right)$ and $S^z_j=n_j-{1\over 2}$, with $n_j=c^\dagger_j c_j$, we can reinterpret $H=H_0+H_I$ as a model for  interacting spinless fermions:
\begin{eqnarray}
H_0&=&\sum_{x;a=1,2}\left({J_a\over 2}c^\dagger_x c_{x+a}+{\rm H.c.}\right)=\sum_k\epsilon(k)c^\dagger_k c_k,\label{eq:H_0}\\
H_I&=&\sum_{x;a=1,2}\left(\Delta_a J_a n_x n_{x+a}\right)-J_2\sum_x\left(c^\dagger_x n_{x+1} c_{x+2}+{\rm H.c.}\right),\label{eq:H_I}.
\end{eqnarray}
Here we have dropped the chemical-potential terms (including $B_z$), which just tune the conserved $\sum_x n_x$. The single-particle dispersion relation is $\epsilon(k)=J_1 \cos (k) +J_2 \cos (2k)$. We assume $J_1<0$ and $|J_1|<4|J_2|$ to guarantee that $\epsilon(k)$ has two minima at $k=\pm Q$ with $\cos (Q)=-{J_1 \over 4 J_2}$ [see Fig.~\ref{fig:1}(b)]. The condition of having a nearly saturated spin chain directly leads to a low-density of fermions, i.e., the dilute limit, in which the Fermi momenta $Q_1, Q_2\rightarrow Q$ [see Fig.~\ref{fig:1}(b)].

To bosonize $H$, we introduce  creation and annihilation operators in the vicinity of the Fermi points: $\psi_a(p)\equiv 
c(Q_a +p)$ and $c(- Q_a +p)\equiv\bar{\psi}_a(p)$ for $a=1,2$. A Fourier-transform of these fields leads to their real space version, 
\begin{equation}
\label{eq:c_exp}
c_x=e^{iQ_1x}\psi_1(x)+e^{-iQ_1x}\bar{\psi}_1(x)+e^{iQ_2x}\psi_2(x)+e^{-iQ_2x}\bar{\psi}_2(x).
\end{equation}
The chiral fields  $\psi_1(x)$ and $\bar{\psi}_1(x)$  vary slowly in space. This is similar to standard bosonization, but with twice the number of species.  After  linearizing the dispersion relation, $\epsilon(\pm Q_1 +p)=\mp v_1 p$ and $\epsilon(\pm Q_2 +p)=\pm v_2 p$ [see Fig. ~\ref{fig:1}(b)], $\psi_2$ and $\bar{\psi}_1$ ($\psi_1$ and $\bar{\psi}_2$) become right (left) movers, and  the chiral fermions can be represented in terms of bosonic fields 
\begin{eqnarray*}
\psi_{1,2}(x)&=&{1\over \sqrt{2\pi}}e^{\pm i\phi_{1,2}(x)},\: \left[\partial_x {\phi}_{1,2}(x),{\phi}_{1,2}(x')\right]=\pm2\pi i \delta (x-x'),\\
\bar{\psi}_{1,2}(x)&=&{1\over \sqrt{2\pi}}e^{\mp i\bar{\phi}_{1,2}(x)},\:
\left[\partial_x \bar{\phi}_{1,2}(x),\bar{\phi}_{1,2}(x')\right]=\mp 2\pi i \delta (x-x').\\
\end{eqnarray*}
The chiral current operators~\cite{note2} can be written as $j_a(x)\equiv \psi^\dagger_a(x)\psi_a(x)={1\over 2\pi} \partial_x \phi_a(x)$ and $\bar{j}_a(x)\equiv \bar{\psi}_a^\dagger(x)\bar{\psi}_a(x)={1\over 2\pi} \partial_x \bar{\phi}_a(x)$.

The noninteracting part of the Hamiltonian density can be written in terms of diagonal chiral current bilinears $j_a (x) j_a (x)$ as $H_0=\pi\sum_{a=1,2}\int dx\left[v_a j_a (x) j_a (x)+v_a \bar{j}_a (x) \bar{j}_a (x)\right]$. The interacting part, which describes various scattering processes, has the  general form:
\begin{equation}\label{eq:eff1}
\begin{split}
H_I=\int dx\Big[&g_{1\bar{1}} j_1(x)\bar{j}_1(x)
+g_{12}j_1(x)j_2(x)+g_{1\bar{2}}j_1(x)\bar{j}_2(x)
\\
&+g_{\bar{1}2}\bar{j}_1(x)j_2(x)
+g_{\bar{1}\bar{2}}\bar{j}_1(x)\bar{j}_2(x)+g_{2\bar{2}}j_2(x)\bar{j}_2(x)\\
&+g_c\left(\psi_1^\dagger(x)\bar{\psi}^\dagger_1(x)\psi_2(x)\bar{\psi}_2(x)+{\rm H.c.}\right)
\Big],
\end{split}
\end{equation}
where the coefficients $g$ represent the effective interactions at the fixed point, where the renormalization-group flow stops. A derivation of the bare coupling constants in terms of the microscopic parameters of the XXZ chain is provided in the Supplemental Material~\cite{suppl}.

We now introduce the  fields
\begin{equation}
\varphi(x)={1\over 2}\left[\phi_1(x)+\phi_2(x)\right],\quad \bar{\varphi}(x)={1\over 2}\left[\bar{\phi}_1(x)+\bar{\phi}_2(x)\right],
\end{equation}
and their conjugate momenta $\Pi(x)=-{1\over 2\pi}\left[\partial_x\phi_1(x)-\partial_x\phi_2(x)\right]$ and $\bar{\Pi}(x)={1\over 2\pi}\left[\partial_x\bar{\phi}_1(x)-\partial_x\bar{\phi}_2(x)\right]$. Physically, $\Pi(x)$ and $\bar{\Pi}(x)$ are proportional to current operators from fermions in the vicinity of  the right and left minimum respectively [see Fig.~\ref{fig:1}(b)]. Similarly, $\partial_x \varphi(x)$ and $\partial_x \bar{\varphi}(x)$ are proportional to densities near these minima.

We are interested in the dilute limit of small (but finite) density of electrons, for which $v_1\approx v_2=v$. When approaching the saturation QCP (zero density), the velocity $v$ vanishes as $Q_1-Q_2$. The momentum cutoff around the Fermi points also decreases proportional to the density. As the renormalized coupling constants continuously approach their value at the QCP, we argue that by approaching saturation, $g_{12}$ and $g_{\bar{1}\bar{2}}$ continuously approach zero as they  are irrelevant at the QCP for precisely the same reason as for the single-minimum case: the Pauli exclusion principle forbids interactions like $\psi^\dagger_x \psi^\dagger_x \psi_x \psi_x$ so the most relevant interactions must have two derivatives, $\psi^\dagger_x \partial_x \psi^\dagger_x \psi_x \partial_x \psi_x$, making them irrelevant perturbations to the free-fermion fixed point (see Ref.~\cite{Sachdev}).  Moreover, the spatial derivative that appears  in the  fermionic currents $i \left(\psi^\dagger_x \partial_x\psi_x-\partial_x\psi^\dagger_x  \psi_x\right)$ makes the coefficient of $\Pi (x)\bar{\Pi}(x)$ irrelevant (the terms proportional to $\Pi^2$ and $\bar{\Pi}^2$ are, however, relevant as the fermionic anticommutation relations yield relevant terms of type $\partial_x \psi^\dagger \partial_x \psi$ for the same species). In addition, inversion symmetry requires $g_{1\bar{2}}=g_{\bar{1}2}$. 

The general form of the Hamiltonian in the dilute limit is then given by
\begin{eqnarray}\label{eq:eff2}
&H &=\left({1 \over 2 \pi}\right)^2\int d x \Big[ 2\pi   v\left[(\partial_x \varphi)^2 + (\partial_x \bar{\varphi})^2\right]
+2\pi^3 v\left(\Pi^2+ \bar{\Pi}^2\right) 
\nonumber \\
&+&g\pi \left(\partial_x \varphi\bar{\Pi} - \partial_x \bar{\varphi}\Pi\right)+
g' \partial_x \varphi\partial_x \bar{\varphi}+2g_c \cos\left[2(\bar{\varphi}-\varphi)\right]
\Big],
\end{eqnarray} 
where  $g\equiv g_{1\bar{1}}-g_{2\bar{2}}$, $g'\equiv g_{1\bar{1}}+2g_{1\bar{2}}+g_{2\bar{2}}$ and the explicit dependence of the fields on $x$ is suppressed. Since we have used the limiting values of the coupling constants in the limit of vanishing density, it is important to bear in mind that our results are valid only over large length scales in comparison with the inter-particle spacing (inverse of the cutoff for linearized dispersion).

If the term proportional to $g_c$ becomes relevant, it can open a gap and destroy criticality. However, we have a quantum liquid if this term is irrelevant (to be checked a posteriori). If  $ g'$ also flows to zero for a certain range of microscopic parameters, we can rewrite the Hamiltonian as
\begin{eqnarray}\label{eq:eff3}
H &=&{ u\over 2\pi}\int d x \sum_{\sigma=\pm}\left[{1\over K}\left(\partial_x \varphi_\sigma\right)^2+K\left(\pi \Pi_\sigma\right)^2\right],
\end{eqnarray}
where the new fields  are related to the old ones through the following anyonic gauge transformation:
\begin{eqnarray*}
\varphi_+&\equiv&\varphi, \quad \Pi_+\equiv\Pi-{\alpha\over \pi^2}\partial_x \bar{\varphi},\quad \varphi_-\equiv\bar{\varphi}, \quad \Pi_-\equiv\bar{\Pi}+{\alpha\over \pi^2}\partial_x {\varphi},
\end{eqnarray*}
with $\alpha\equiv{g \over 4 v}$, $K=1/\sqrt{1-\left(\alpha\over \pi\right)^2}$, and $u=v/K$~\cite{note3}. Note that the momentum of one species is shifted by a gauge field times the density of the other species. This is equivalent to attaching a  flux to each particle in such a way that the new ``composite'' particles  obey anyonic commutation relations~\cite{Schulz98}: $\alpha$ represents the mutual statistical phase for exchanging the two types of particles. In other words, the anyonic nature of the new quasiparticles corresponds to a generalized J-W transformation (discussed below) and can be inferred from the commutation relations given below Eq.~\eqref{eq:c_exp}~\cite{Schulz98}. Because the scaling dimension of $\cos\left[2(\bar{\varphi}-\varphi)\right]$ is $2K$ for the anyonic liquid, $g_c$ indeed flows to zero. 

In fact, the Hamiltonian~\eqref{eq:eff3} is a direct generalization of the Shastry-Schulz model of noninteracting anyons \cite{Schulz98}.   Just like in the  Shastry-Schulz model,   the two anyonic species are completely decoupled  (there is a unique statistics of quasiparticles for which the theory breaks into two decoupled sectors).  The Shastry-Schulz model, however,  corresponds to the special case of $K=1$, indicating no intra-species interactions. The $\alpha$-dependent $K$ in our model results in a continuous interpolation from free bosons ($\alpha \to \pi$, $H={ \pi v\over 2}\int d x \sum_{\sigma}\Pi_\sigma^2$ ) to free fermions ($\alpha=0$, $K=1$).

The key to realizing the anyonic liquid \eqref{eq:eff3}, however, is a vanishing renormalized $g'$ at the fixed point. Although it is difficult to express $g'$ in terms of microscopic parameters, an exact two-magnon calculation allows us to determine the microscopic parameters for which $g'=0$. We use the analogy with free fermions (a LL with Luttinger parameter $K=1$). For such noninteracting LL, the  two-particle state $c^\dagger_{k_1}c^\dagger_{k_2} |0\rangle$ is an exact eigenstate of the Hamiltonian. As soon as $K$ moves away from unity, this state scatters into other  two-particle states and will not remain an eigenstate. Thus, if the effective Hamiltonian has the general Luttinger-liquid form and $c^\dagger_{k_1}c^\dagger_{k_2} |0\rangle$ is an exact eigenstate of the microscopic Hamiltonian, the Luttinger parameter must be equal to unity (free-fermion fixed point). Similarly, we require that a two-anyon state is an exact eigenstate of the Hamiltonian \eqref{eq:xxz}.

Going back to Eq.~\eqref{eq:xxz}, we perform a generalized J-W transformation to anyons with statistical phase $\phi$ and annihilation operator $a_x$ on site $x$: $S^{-}_{x} = a_x e^{-i \phi \sum_{y<x} n_{y}}$ and $S^z_{x} = n_x -\frac{1}{2}$ with $n_x=a^\dagger_x a_x$. The anyonic statistics of these particles can be observed in the relationship $a^\dagger_x a^\dagger_y=e^{-i\phi}a^\dagger_y a^\dagger_x$ for $x<y$ (see Ref.~\cite{Batista12} for the physical interpretation of anyons in terms of spins). In the dilute limit, the possible momenta are $\pm Q$. We need to find a relationship between the microscopic parameters so that the two-particle state $a^\dagger_Q a^\dagger_{\bar {Q}}|0\rangle$, with $\bar {Q}\equiv -Q$, where $a_Q$ is the Fourier transform of $a_x$ defined above at momentum $Q$, is an exact eigenstate of Eq.~\eqref{eq:xxz}.  The Hamiltonian has the same form as Eqs.~\eqref{eq:H_0} and  \eqref{eq:H_I} in terms of anyonic operators (with $c$ replaced by $a$), except for the correlated hopping term (the term in $H_I$ proportional $J_2$), which now reads $\frac{J_2}{2} \sum_x n_{x+1} \left[ (e^{i\phi} -1) a^{\dagger}_{x} a^{\;}_{j+2} + (e^{-i\phi} -1) a^{\dagger}_{x+2} a^{\;}_{x} \right]$. Requiring $H a^\dagger_Q a^\dagger_{\bar {Q}}|0\rangle=\epsilon a^\dagger_Q a^\dagger_{\bar {Q}}|0\rangle$ leads to 
\begin{eqnarray}
\Delta_1&=& \cos (Q) +\frac{\sin{(Q)}} {2} \left [  \tan{(Q)} +  \tan{(Q+\phi/2)}  \right ],\label{eq:rel1}
 \\
\Delta_2 &=& \cos{(2Q)}  + \sin{(2Q)} \tan{(2Q+\phi/2)},\label{eq:rel2}
\end{eqnarray}
with the energy given by $\epsilon = -2 (\Delta_1 J_1 + \Delta_2 J_2) + 2 J_1 \cos{(Q)} + 2 J_2 \cos{(2Q)}$. Note that eliminating $\phi$ between Eqs.~\eqref{eq:rel1} and \eqref{eq:rel2} gives a relationship between the microscopic parameters $\Delta_1$ and $\Delta_2$ for a given $J_1/J_2$ ($\cos Q=-J_1/4J_2$). This relationship is achieved by tuning only one microscopic parameter and it allows the system to realize an anyonic liquid with an emergent statistical angle $\phi$  determined by the above equations.  
Because there is only one anyon of each species in $a^\dagger_Q a^\dagger_{\bar {Q}}|0\rangle$, the intra-species interactions characterized by the parameter $K$, play no role in the above argument. 
\begin{figure}
\vspace{0.3cm}
 \includegraphics[width =8cm]{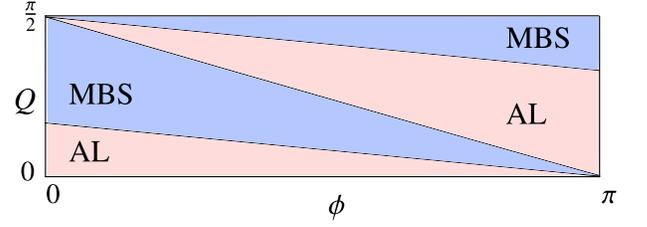}
\caption{The phase diagram of the Hamiltonian~\eqref{eq:xxz} with ${J_1 \over 4 J_2}=-\cos(Q)$ and other coupling constants given by Eqs.~\eqref{eq:rel1} and \eqref{eq:rel2}. The phases are respectively denoted by AL (anyonic liquid) and MBS (magnon bound state).}
 \label{fig:2}
\end{figure}

If the effective theory of the system is given by Eq.~\eqref{eq:eff2}, the above values of $\Delta_1$ and $\Delta_2$ guarantee the absence of scattering between the two anyonic species. The effective Hamiltonian must then reduce to  Eq.~\eqref{eq:eff3} with $\alpha= \pi-\phi$. In other words, we have a family of Hamiltonians characterized by two parameters $Q$ and $\phi$, which can potentially flow to the anyonic-liquid fixed point \eqref{eq:eff3}. However, the formation of low-energy bound states may lead to either a first-order phase transition from the saturated state (the number of particles changes discontinuously at the saturation field) or a continuous transition into a state with dominant nematic (BEC of pairs) or higher-order multipolar fluctuations. As discussed in the supplemental material~\cite{suppl}, by using exact two-magnon calculations~\cite{Gochev1974}, we found the range of parameters that give rise to low-energy bound states, destabilizing the anyonic liquid, and obtained the phase diagram of Fig.~\ref{fig:2}.

Returning to the anyonic liquid, we now present analytical predictions for different correlation functions, which are numerically verified with the DMRG method. For the fermionic Green's function $G(x)=\langle c^\dagger_y c_{x+y}\rangle$, we find 
\begin{equation}\label{ferm}
G(x)\propto \left[\sin\left(Q_1 x +\omega_1\right)-\sin\left(Q_2 x +\omega_2\right)\right]x^{  -{1/ \sqrt{1-\lambda^2}} },
\end{equation}
for $x \rho_0 \gg 1$, with $\lambda=\alpha/\pi$ in the dilute limit. In general, the ordering vectors change at finite densities (because of the string operator that relates fermions to anyons), but the change is negligible in the limit of small density considered here~\cite{Schulz98}. Moreover, the ordering vectors $Q_i$ have an uncertainty of order $1\over L$ in a finite system of length $L$. We therefore compare the above prediction with the numerical results by fitting the numerically computed correlation function to expression~\eqref{ferm}, with the ordering vectors, the overall coefficient, and the exponent as fitting parameters (using the fact that the exponents are relatively close to 1 we neglect the phase shifts in the oscillatory prefactor \cite{suppl} in fitting the data). An exponent close to $-{1\over \sqrt{1-\lambda^2}}  $ and ordering vectors close to the computed  (for the given density of fermions) $Q_1$ and $Q_2$ would corroborate our analytical prediction for an anyonic liquid.
\begin{figure}
 \includegraphics[width =8 cm]{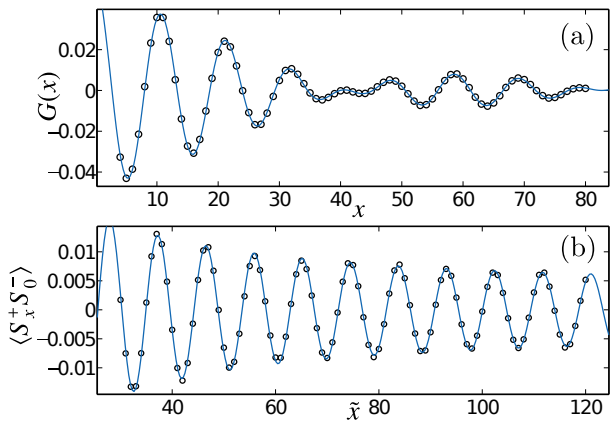}
\caption{(a) The fermionic Green's function for $\phi/\pi=0.615$ and $Q/\pi=0.2$ at density $\rho_0=0.05$. The black circles (blue line) represent numerical results (fit). Fitting to Eq.~\eqref{ferm} gives $Q_1/\pi=0.16$, $Q_2/\pi=0.21$ and an exponent $0.108$ in excellent agreement with analytical predictions $Q_1/\pi=0.17$, $Q_2/\pi=0.22$ and an exponent $0.108$. (b) The spin-spin correlation function for the same parameter. Fitting to Eq.~\eqref{spin_corr} gives an exponent $0.70$ in good agreement with the analytical prediction $0.67$.}
 \label{fig:3}
\end{figure}

We performed the DMRG calculations for a chain of length $L=400$ with periodic boundary conditions (implemented by constructing two parallel chains of length $L/2$ and connecting  the endpoints~\cite{Rahmani}). We  compared the results with a calculation for  $L=200$ and chose the range of $x$ where the two data sets overlap. Excellent convergence was obtained by keeping 1000 states in the DMRG iterations. As seen in Fig.~\ref{fig:3}(a), the exponent of the correlation function differs from $\delta=1$  (free fermion fixed point) and it is consistent with the exponents of an anyonic liquid. The ordering momenta are also very close to our analytical predictions (the agreement cannot be perfect because of the finite value of the density $\rho_0=0.05$).

These statistical angle also changes the asymptotic bahavior of the two-point spin-spin correlators. This angle can then be  obtained by measuring the the $k$-dependence of the transverse  magnetic  susceptibility $\chi_{xx}=\chi_{yy}$, which is determined by the Fourier transform of the correlator $\langle S^+_x S^-_0\rangle$.  At low densities, we can neglect the average density $\rho_0$ in $\sum_{y<x} n_y=\int_{-\infty}^x dy \left[\rho_0 + \sum_a\left( j_a(y)+\bar{j}_a(y)\right)\right]$ and write $S^-_x \sim c_x e^{-i\left[\varphi(x) +\bar{\varphi}(x)\right]}$. By using Eqs.~\eqref{eq:c_exp} and \eqref{eq:eff3}, we find that the four terms in the $\langle S^+_x S^-_0\rangle$ fall into two categories, respectively decaying to leading order as $x^{-{1\over 2}\left(\sqrt{1-\lambda^2}+\left(1\pm \lambda\right)^2/\sqrt{1-\lambda^2}\right)}$ (where $\lambda={\alpha\over \pi}$), with the leading dilute-limit behavior given by
\begin{equation}\label{spin_corr}
\langle S^+_x S^-_0\rangle\propto \sin\left(Q_1x+\omega\right){x^{-{1\over 2}\left(\sqrt{1-\lambda^2}+\left(1- \lambda\right)^2/\sqrt{1-\lambda^2}\right)}},
\end{equation}
for $x \rho_0 \gg 1$ ($\omega$ is a phase shift). We also checked the above expression with DMRG. The bosonic correlators have a stronger finite-size dependence so in fitting the data we replaced $x$ in ${x^{-{1\over 2}\left(\sqrt{1-\lambda^2}+\left(1- \lambda\right)^2/\sqrt{1-\lambda^2}\right)}}$ with its finite-size counterpart $\tilde{x}={L \over \pi}\sin \left(\pi {x\over L}\right)$. The agreement is excellent as shown in Fig.~\ref{fig:3}(b).
The anyonic fixed point can be detected by comparing the above exponent  with the exponent of the correlator that determines the {\it longitudinal} susceptibility $\chi_{zz}$: the oscillatory [$k=\pm(Q_2 - Q_1)$] components  of $\langle S^z_x S^z_0\rangle$ decay as $x^{-{1\over 2K}}=x^{-{1\over 2}\sqrt{1-\lambda^2}}$ { for $x \rho_0 \gg 1$~\cite{note4}. Finally, we note that disorder is a relevant perturbation for magnetic saturation  QCP's~\cite{note5}. However, the exponents  that we are predicting for the two-spin  correlators can still be measured if the  characteristic length scale associated with the  disorder is much longer than the average inter-particle distance $1/\rho_0$.

In summary, by studying the effects of strong magnetic frustration in nearly saturated spin chains, we extended the classification of the saturation QCPs  from the standard paradigm of simple free fermionic (bosonic) theories in $d=1$ ($d>1$)~\cite{Sachdev} to an exotic continuous line of anyonic liquids,. These liquids are characterized by two species of anyonic quasiparticles with vanishing inter-species interactions. The emergent statistical phase of the quasiparicles interpolates continuously between bosons and fermions. While envisioned in the field-theory literature, anyonic liquids had thus far remained as an abstract theoretical construction. Our results provide natural realizations of one-dimensional anyonic liquids in a  simple and  experimentally relevant model, opening a promising direction in the search for anyons in frustrated magnets. As only one exchange parameter needs to be tuned in order to realize our anyonic liquids (apart from the magnetic field which can be easily brought to the vicinity of the critical point), physical or chemical pressure could drive generic highly frustrated one-dimensional magnetic materials into the anyonic-liquid phase. Relationships between the transverse and longitudinal magnetic susceptibilities serve as experimental signatures of this exotic phase. The fate of higher-dimensional systems realized by coupling these anyonic wires~\cite{Kane02, teo,oreg,Mong} poses an interesting challenge for future investigations. For certain anyonic phases~\cite{note6}, novel two-dimensional topological phases might emerge (see Refs. \cite{Mansson,Lahtinen} for such constructions).

\begin{acknowledgements}
We are grateful to Ian Affleck, Claudio Chamon, Leonid Glazman, and Yong Baek Kim for helpful discussions. We acknowledge support by the U.S. DOE through LANL/LDRD program (A.R. and C.D.B) and NSF through grant DMR-1339564 (A.E.F).
\end{acknowledgements}


\begin{thebibliography}{99}
\bibitem{note1}
While exchange statistics is not well defined in one dimension because a transmutation of the statistics can be absorbed in the interactions, it is still meaningful to single out a unique statistics that makes the  quasiparticles weakly interacting.
\bibitem{note2}
Although commonly referred to as currents, these are physically the densities of the chiral fermions.
\bibitem{note3}
This is a more general $(c=1)\times(c=1)$ conformal field theory than the standard spinful Luttinger liquid as it allows for more marginal terms in the Hamiltonian. See Ref. \cite{Pham00} for a detailed discussion.
\bibitem{note4}The leading non-oscillatory component decays as $x^{-2}$ but extracting the oscillatory part is not too difficult.
\bibitem{note6}The continuously tunable anyonic phase does not seem to allow for the emergence of rigid topologically ordered phase for all anyonic phases.
\end{thebibliography}
\end{document}

\title{Supplemental Material for ``Anyonic Liquids in Nearly Saturated Spin Chains''}

\author{Armin Rahmani}
\affiliation{
Theoretical Division, T-4 and CNLS, Los Alamos National Laboratory, Los Alamos, New Mexico 87545, USA} 

\author{Adrian Feiguin}
\affiliation{
Department of Physics, Northeastern University, Boston, Massachusetts 02115, USA}

\author{Cristian D. Batista}
\affiliation{
Theoretical Division, T-4 and CNLS, Los Alamos National Laboratory, Los Alamos, New Mexico 87545, USA} 

\date{\today}
\pacs{}

\maketitle

\section{Bare coupling constants}
If we neglect the variations of the slow fields over a distance of order a few lattice spacings, inserting Eq.~(4) of the main text into the expression for $H_I$ in Eq.~(3)  leads to the form given in Eq.~(5)  with the bare coupling constants below:
\begin{eqnarray}
\tilde{g}_{1\bar{1}}&=&4\Delta_1J_1 \sin^2 (Q_1)+4 \Delta_2J_2 \sin^2 (2Q_1)+8J_2\sin^2  (Q_1),\\
\tilde{g}_c&=&-4\Delta_1J_1 \sin (Q_1) \sin (Q_2)-4 \Delta_2J_2 \sin (2Q_1)\sin (2Q_2)-8J_2\sin (Q_1)\sin (Q_2),\\
\tilde{g}_{12}&=&4\Delta_1J_1 \sin^2 \left({Q_1-Q_2\over 2}\right)+4 \Delta_2J_2 \sin^2 (Q_1-Q_2)+2J_2\left[2\cos(Q_1+Q_2)-\cos(2Q_1)-\cos(2Q_2)\right],\\
\tilde{g}_{1\bar{2}}&=&4\Delta_1J_1 \sin^2 \left({Q_1+Q_2\over 2}\right)+4 \Delta_2J_2 \sin^2 (Q_1+Q_2)+2J_2\left[2\cos(Q_1-Q_2)-\cos(2Q_1)-\cos(2Q_2)\right],\\
\tilde{g}_{\bar{1}2}&=&4\Delta_1J_1 \sin^2 \left({Q_1+Q_2\over 2}\right)+4 \Delta_2J_2 \sin^2 (Q_1+Q_2)+2J_2\left[2\cos(Q_1-Q_2)-\cos(2Q_1)-\cos(2Q_2)\right],\\
\tilde{g}_{\bar{1}\bar{2}}&=&4\Delta_1J_1 \sin^2 \left({Q_1-Q_2\over 2}\right)+4 \Delta_2J_2 \sin^2 (Q_1-Q_2)+2J_2\left[2\cos(Q_1+Q_2)-\cos(2Q_1)-\cos(2Q_2)\right],\\
\tilde{g}_{\bar{2}\bar{2}}&=&4\Delta_1J_1 \sin^2 (Q_2)+4 \Delta_2J_2 \sin^2 (2Q_2)+8J_2\sin^2  (Q_2).
\end{eqnarray}
The above expressions for the coupling constants in terms of the microscopic parameters are only valid in the limit of small interactions $|H_I|\ll |H_0|$. Generally, we can not neglect the short-distance (high-energy) physics stemming from the variations of the slow fields. Integrating them out, however, does not change the form of $H_I$ (as it includes all allowed scattering processes); it merely renormalizes the coupling constants.

To derive the expression above, it is convenient to define $H_{aba'b'}\equiv\int dx c^\dagger_{x+a} c^\dagger_{x+b}c_{x+a'}c_{x+b'}$. Now, each of the creation and annihilation operators in $H_{aba'b'}$ can be written as a linear combination of four chiral operators as in Eq. (4) of the main text. As mentioned above, $a$, $b$, $a'$, and $b'$ are assumed of the order the lattice spacing. The chiral fields have slow variations over such distances and we have used, e.g., $\psi_j(x+a)\approx \psi_j(x)$ and $\bar{\psi}_j(x+a)\approx \bar{\psi}_j(x)$. Assuming the momenta $Q_j$ do not take any special values that allow for Umklapp processes, all terms (out of the $4^4$ terms coming from expanding the product in the integrand) that have an $x$-dependent oscillatory factor vanish upon integration due to momentum conservation. We can then write
\begin{equation}
\begin{split}
H_{aba'b'}\approx \int dx \Big[
&4\sin\left[Q_1(a-b)\right]\sin\left[Q_1(a'-b')\right]
{\psi}^\dagger_1\bar{\psi}^\dagger_1{\psi}_1\bar{\psi}_1\\
+
&4\sin\left[Q_1(a-b)\right]\sin\left[Q_2(a'-b')\right]
{\psi}^\dagger_1\bar{\psi}^\dagger_1{\psi}_2\bar{\psi}_2\\
+
&\left(
e^{iQ_1(a'-a)+iQ_2(b'-b)}
-e^{iQ_1(b'-a)+iQ_2(a'-b)}
-e^{iQ_1(a'-b)+iQ_2(b'-a)}
+e^{iQ_1(b'-b)+iQ_2(a'-a)}
\right)
{\psi}^\dagger_1\psi^\dagger_2\psi_1{\psi}_2\\
+
&\left(
e^{iQ_1(a'-a)-iQ_2(b'-b)}
-e^{iQ_1(b'-a)-iQ_2(a'-b)}
-e^{iQ_1(a'-b)-iQ_2(b'-a)}
+e^{iQ_1(b'-b)-iQ_2(a'-a)}
\right)
{\psi}^\dagger_1\bar{\psi}^\dagger_2\psi_1\bar{\psi}_2\\
+
&\left(
e^{iQ_1(a-b)-iQ_2(b+a')}
-e^{iQ_1(a-a')-iQ_2(b+b')}
-e^{iQ_1(b-b')-iQ_2(a-a')}
+e^{iQ_1(b-a')-iQ_2(a-b')}
\right)
\bar{\psi}^\dagger_1\psi^\dagger_2\psi_2\bar{\psi}_1\\
+
&\left(
e^{iQ_1(a-a')+iQ_2(b-b')}
-e^{iQ_1(a-b')+iQ_2(b-a')}
-e^{iQ_1(b-a')+iQ_2(a-b')}
+e^{iQ_1(b-b')+iQ_2(a-a')}
\right)
\bar{\psi}^\dagger_1\bar{\psi}^\dagger_2\bar{\psi}_1\bar{\psi}_2\\
+
&4\sin\left[Q_1(a'-b')\right]\sin\left[Q_2(a-b)\right]
{\psi}^\dagger_2\bar{\psi}^\dagger_2{\psi}_1\bar{\psi}_1\\
+
&4\sin\left[Q_2(a'-b')\right]\sin\left[Q_2(a-b)\right]
{\psi}^\dagger_2\bar{\psi}^\dagger_2{\psi}_2\bar{\psi}_2
\Big],
\end{split}
\end {equation}
where all the chiral fields are calculated at position $x$. The dependence of $\tilde{g}$ on the microscopic parameters then readily follows from the relationship $H_I=-\Delta_1 J_1 H_{0101}-\Delta_2 J_2 H_{0202}-J_2 \left(H_{0112}+H_{2110}\right),
$

The most general Hamiltonian before invoking inversion symmetry and the irrelevance of several coupling constants in the dilute limit can be written as
\begin{equation}
\begin{split}
H=\left({1 \over 2 \pi}\right)^2\int d x \Big\{&\left(g_{12}+\pi v_1+\pi v_2\right)\left[\partial_x \varphi(x)\right]^2+
\left(g_{\bar{1}\bar{2}}+\pi v_1+\pi v_2\right)\left[\partial_x \bar{\varphi}(x)\right]^2\\
&+\pi^2(-g_{12}++\pi v_1+\pi v_2)\left[\Pi(x)\right]^2+\pi^2(-g_{\bar{1}\bar{2}}+\pi v_1+\pi v_2)\left[\bar{\Pi}(x)\right]^2\\
&+\pi (g_{12}+\pi v_1-\pi v_2)\Pi(x)\partial_x \varphi(x)+\pi (-g_{12}+\pi v_1-\pi v_2)\left[\partial_x \varphi(x)\right]\Pi(x)\\
&-\pi (g_{\bar{1}\bar{2}}+\pi v_1-\pi v_2)\bar{\Pi}(x)\partial_x \bar{\varphi}(x)-\pi (-g_{\bar{1}\bar{2}}+\pi v_1-\pi v_2)\left[\partial_x \bar{\varphi}(x)\right]\bar{\Pi}(x)\\
& +(g_{1\bar{1}}+g_{1\bar{2}}+g_{\bar{1}2}+g_{2\bar{2}})\left[\partial_x \varphi(x)\right]\left[\partial_x \bar{\varphi}(x)\right]\\
&+\pi^2 (-g_{1\bar{1}}+g_{1\bar{2}}+g_{\bar{1}2}-g_{2\bar{2}})\Pi (x)\bar{\Pi}(x)\\
&+\pi(-g_{1\bar{1}}+g_{1\bar{2}}-g_{\bar{1}2}+g_{2\bar{2}})\partial_x \varphi(x)\bar{\Pi}(x)\\
&+\pi(g_{1\bar{1}}+g_{1\bar{2}}-g_{\bar{1}2}-g_{2\bar{2}})\partial_x \bar{\varphi}(x)\Pi(x)\\
&+2g_c \cos\left[\bar{\varphi}(x)-\varphi(x)\right]
\Big\}.
\end{split}
\end{equation}
Notice that the bare values of the coupling constants, which are irrelevant in  the dilute limit, vanish as $(Q_1-Q_2)^2$.

\section{Magnon bound states}
The bound states are most easily analyzed in the original spin representation. 
%We denote the fully saturated state $|\uparrow \uparrow \uparrow \dots \rangle$  by $|0\rangle$. The states $|\phi_{i,j}\rangle=S^-_i S^-_j |0\rangle$ with $i<j$ provide a basis for the  the two-magnon subspace. Translation invariance implies that an arbitrary two-magnon eigenstate has a well defined center-of-mass momentum $q$: $|\Psi\rangle=\sum_{i,j>i} e^{iq R_{i,j}} u(r_{i,j})|\phi_{i,j}\rangle$ for $R_{i,j}=\left(r_i+r_j\right)/2$ and $r_{i,j}=r_j-r_i$. The eigenvalue equation $H|\psi\rangle=\epsilon|\psi\rangle$ then gives coupled equations for $u(r)$~\cite{suppl}.  
We denote the vacuum $|\uparrow \uparrow \uparrow \dots \rangle$ by $|0\rangle$ and represent the two-magnon states as
\begin{equation}
|\phi_{i,j}\rangle=S^-_i S^-_j |0\rangle, \qquad i<j.
\end{equation}
Due to translation invariance of the Hamiltonian, two-magnon eigenstates have a well-defined center-of-mass momentum $q$:
\begin{equation}
|\psi\rangle=\sum_{i,j>i} e^{iq R_{i,j}} u(r_{i,j})|\phi_{i,j}\rangle,\qquad R_{i,j}=\left(r_i+r_j\right)/2,\qquad r_{i,j}=r_j-r_i.
\end{equation}
The eigenvalue equation $H|\psi\rangle=\epsilon|\psi\rangle$ in this sector [with $H$ given by Eq. (1) of the main text] then reduces to
\begin{eqnarray}
\left(\epsilon +J_1^z+2J_2^z \right) u(1)&=&J_2 \cos\left(q\right)u(1)+ J_1 \cos\left({q\over 2}\right) u(2)+ J_2 \cos\left(q\right)u(3),\label{eq:rel1}\\
\left(\epsilon+2J_1^z+J^z_2\right) u(2)&=&J_1  \cos\left({q\over 2}\right)u(1)+ J_1 \cos\left({q\over 2}\right) u(3)+ J_2 \cos\left(q\right)u(4),\label{eq:rel2}\\
\left(\epsilon+2J_1^z+2J^z_2\right) u(r)&=& J_1 \cos\left({q\over 2}\right) \left[u(r-1)+u(r+1)\right]\label{eq:rel3}\\ \nonumber
& &+ J_2 \cos\left(q\right)\left[u(r-2)+u(r+2)\right],\qquad r>2.
\end{eqnarray}
The last relationship above for $r>2$ [Eq.~\eqref{eq:rel3}] (in the bulk) has exponential solutions $u(r)=e^{-\kappa r}$, where $z=e^{-\kappa}$, for a $\kappa$ on the complex plane, satisfies the characteristic polynomial equation $\epsilon+2J_1^z+2J^z_2= J_1 \cos\left({q\over 2}\right) \left(z+{1\over z}\right)+ J_2 \cos\left(q\right)\left(z^2+{1\over z^2}\right)$.
We generally obtain a continuum of plane-wave scattering solutions giving rise to a continuous spectrum, but it is also possible to obtain bound states, which correspond either to a single exponentially decaying solution $u(r)=e^{-\gamma r}$ for $r\geqslant 2$ (for a real positive $\gamma$) or a linear combination of two such solutions $u(r)=e^{-\kappa r}+e^{i\theta}e^{-\kappa^* r}$ with ${\rm Re}({\kappa})>0$ for $r\geqslant 1$ (where $\theta$ is a phase shift).

 For any energy, there are four solutions for $z$ but solutions with ${\rm Re}(\kappa)<0$ ($|z|>1$) are unphysical as they can not be normalized.  Wave functions with $|z|=1$ are extended scattering states, while wave functions with $|z|<1$ are bound states. Since for a given solution $z$, $1\over z$ and $z^*$ are also solutions to the characteristic equation, there are at most two independent bound-state solutions with $|z|<1$. An ansatz bound-state solution satisfying the boundary conditions \eqref{eq:rel1} and \eqref{eq:rel2} is then given by a linear combination of these normalizable wave functions: $u(r)=e^{-\kappa r}+ s e^{-\kappa^* r}$ for all $r$, where ${\rm Im}(\kappa)\neq 0$. To satisfy Eqs.~\eqref{eq:rel1} and \eqref{eq:rel2}, we then need $s=-\Upsilon/\Upsilon^*=-\Xi/\Xi^*$ (therefore $|s|=1$), where
\begin{eqnarray}
\Upsilon&\equiv&\equiv-\left(\epsilon +J_1^z+2J_2^z \right) e^{-\kappa }+J_2 \cos\left(q\right)e^{-\kappa }+ J_1 \cos\left({q\over 2}\right) e^{-2\kappa }+ J_2 \cos\left(q\right) e^{-3\kappa },\\
\Xi&\equiv&-\left(\epsilon+2J_1^z+J^z_2\right) e^{-2\kappa }+J_1  \cos\left({q\over 2}\right)e^{-\kappa }+ J_1 \cos\left({q\over 2}\right) e^{-3\kappa }+ J_2 \cos\left(q\right)e^{-4\kappa }.
\end{eqnarray}
The condition for this ansatz is then ${\rm Im}(\Upsilon^*\Xi)=0$. We then scan all energies below the minimum of the two-particle continuum and above exact lower bounds for the two-magnon energy, find $\kappa$ by solving the characteristic equation, and check the condition  ${\rm Im}(\Upsilon^*\Xi)=0$.

Another possibility is that there are real solutions for $z$ and a single exponential satisfies the equations. In this case, we can not require $u(1)$ to have the same form $u(r)=e^{-\gamma r}$ for $r\geqslant 2$. However, we can simply eliminate $u(1)$ and obtain the condition
\begin{equation}
\begin{split}
&\left[2 J_1 \cos\left({q\over 2}\right) \cosh\left(\gamma\right)+2 J_2 \cos\left(q\right)\cosh\left(2\gamma\right)-J_1^z-J_2 
\cos\left(q\right)\right]\\
&\times \left[2 J_1 \cos\left({q\over 2}\right) \cosh\left(\gamma\right)+2 J_2 \cos\left(q\right)\cosh\left(2\gamma\right)-J_2^z-J_1 
\cos\left({q\over 2}\right)e^{-\gamma}\right]\\
&=J_1 \cos\left({q\over 2}\right)\left[J_1\cos\left({q\over 2}\right)+J_2 
\cos\left({q}\right)e^{-\gamma} \right].
\end{split}
\end{equation}

If such solutions exist for some center-of-mass momentum $q$, and the corresponding energy is below the two-particle continuum, the system will form low-energy bound states in the two-magnon sector and it is vulnerable to phase separation. 
Checking for the two types of bound states above, we obtain the phase diagram shown in Fig.~2 of the main text. We have also checked this phase diagram by direct numerical calculation of the ground-state energy with Lanczos diagonalization in the two-particle subspace in a finite system of $L=100$, which showed excellent agreement.

\section{Scaling dimensions}
The correlation functions presented in the main text can be computed easily from Eq. (4) of the main text using the mapping of the chiral modes to new chiral modes that give rise to two noninteracting Luttinger liquids:
\begin{eqnarray}
\phi'_1+\phi'_2&=&{1\over \sqrt{K}}\left(\phi_1+\phi_2\right),\qquad {\phi}'_2-{\phi}'_1={ \sqrt{K}}\left({\phi}_2-{\phi}_1-\lambda\bar{\phi}_1-\lambda\bar{\phi}_2\right),
\\
\bar{\phi}'_1+\bar{\phi}'_2&=&{1\over \sqrt{K}}\left(\bar{\phi}_1+\bar{\phi}_2\right),
\qquad
\bar{\phi}'_1-\bar{\phi}'_2={ \sqrt{K}}\left(\bar{\phi}_1-\bar{\phi}_2+\lambda{\phi}_1+\lambda{\phi}_2\right),
\end{eqnarray}
which gives
\begin{equation}
\left(\begin{array}{c}
\phi_1 \\ 
\bar{\phi}_1\\ 
\phi_2\\ 
\bar{\phi}_2
\end{array} \right)=
{1\over 2}
\left(
\begin{array}{cccc}
{K+1\over \sqrt{K}} & -\lambda\sqrt{K} & {K-1\over \sqrt{K}}  & -\lambda\sqrt{K} \\ 
-\lambda\sqrt{K} & {K+1\over \sqrt{K}} & -\lambda\sqrt{K} &  {K-1\over \sqrt{K}}  \\ 
 {K-1\over \sqrt{K}}  & \lambda\sqrt{K} & {K+1\over \sqrt{K}} & \lambda\sqrt{K} \\ 
\lambda\sqrt{K} &  {K-1\over \sqrt{K}}  & \lambda\sqrt{K}  & {K+1\over \sqrt{K}}
\end{array} 
\right)
\left(\begin{array}{c}
\phi'_1 \\ 
\bar{\phi}'_1\\ 
\phi'_2\\ 
\bar{\phi}'_2
\end{array} \right).
\end{equation}
As the free-fermion chiral correlation functions are known, all correlators of vertex operators can be easily computed from the above expression. For example, $\langle e^{-i\phi_1(0)}e^{i\phi_1(x)}\rangle$, which appears in $G(x)$ is given by
\begin{equation}
\langle e^{-i\phi_1(0)}e^{i\phi_1(x)}\rangle=
\left({i\over x}\right)^{{1\over 4K}\left({K+1}\right)^2}
\left(-{i\over x}\right)^{{1\over 4}\lambda^2 K}
\left(-{i\over x}\right)^{{1\over 4K}\left({K-1}\right)^2}
\left({i\over x}\right)^{{1\over 4}\lambda^2 K},
\end{equation}
which leads to Eq.~(11) of the main text. In case of the spin-spin correlation functions, the exponents are far from unity and the phase shifts $\omega$ can not be neglected.